\documentstyle[11pt,newpasp,twoside,epsf]{article}
\def\edcomment#1{\iffalse\marginpar{\raggedright\sl#1\/}\else\relax\fi}
\reversemarginpar

\begin{document}

\title{A self-regulating braking mechanism in black hole X-ray binaries}
\author{Friedrich Meyer}
\author{Emmi Meyer-Hofmeister}
\affil{Max-Planck-Institut f\"ur Astrophysik, Karl
Schwarzschildstr.~1, D-85740 Garching, Germany}

\begin{abstract}
The outbursts of black hole X-ray transients can be understood as
caused by a limit cycle instability in the accretion disk, similar to
dwarf nova outbursts. For adequately low mass overflow rates from
the companion star long outburst recurrence times are expected. But
the fact that we find predominantly long recurrence times or that only
one X-ray nova outburst was detected at all poses a problem. The
question arises whether any braking mechanism could act in a way that long
recurrence times are favored.

We suggest that a circumbinary disk exists and brakes the orbital
motion of the binary stars by tidal interaction.
The irradiation during an outburst leads to mass loss by winds
from the circumbinary disk, releaving the braking
force until the removed matter is refilled by diffusion from outer
parts. We show that this reduction of braking will self-adjust 
the mass transfer to the marginal rate that gives long recurrence
times.
\end{abstract}

\section{Introduction}
Soft X-ray transients (SXT) are binaries where a low-mass secondary
star transfers mass by Roche lobe
overflow to the compact primary via an accretion disk. The primary can
either be a neutron star or a black hole. For
black hole binaries the outbursts mostly occur after extremely long
quiescence intervals. These so-called X-ray novae are usually detected in
X-rays during the outburst. But for several systems only one outburst
is known, that means the recurrence time is longer than at least 30,
maybe 50 years. A recent review by Charles (1998) on black holes in
our galaxy gives a comprehensive description of the observations.

The evolution of the accretion disk and the triggering of an outburst
can be understood and modelled in the same way as dwarf nova
outbursts. The first investigations were made  by Huang \& Wheeler (1989)
and Mineshige \$ Wheeler (1989). Later detailed investigations by
Cannizzo et al. (1995) and Cannizzo (1998, review 1999) including evaporation
(in a simplified way) and the effects of irradiation
focussed on modelling the exponential decline of the
outburst. For the evolution of the disk during quiescence
one feature, present also in dwarf nova disks but of minor role, is
important in SXTs. It is
necessary to have a truncated inner disk,
otherwise the disk cannot be cool throughout. Such a geometrically
thin cool disk results naturally from evaporation of the innermost disk region
(Meyer \& Meyer-Hofmeister 1994) 
into a corona/ADAF. Using the advection-dominated flow (ADAF) a number of  
quiescent black hole soft X-ray transients, A0620-00, V404 Cyg
(Narayan et al. 1996, 1997), Nova Muscae (Esin et al. 1997) and
GRO J1655-40 (Hameury et al. 1997) have been modelled successfully.
The ADAF in the inner disk
region together with the geometrically thin disk further out, the
transition produced by evaporation, provides a consistent picture for the
accretion disk in SXTs and allows to understand the outburst cycles. 
We computed the disk evolution consistently (as will be
described later) and found a minimum mass transfer
rate to trigger the instability. An important ingredient is that
matter flows through the disk into the corona continuously.

A problem is posed by the very long recurrence times. observed. As the 
modelling of SXT disk evolution has shown (Meyer-Hofmeister \& Meyer 2000)
a spread in the rates of mass overflow from the companion will result
in a spread of recurrence times. That means both, binaries with rare
outbursts and systems with more frequent oubursts, 
should be found. But the observations indicate many more cases of
extremely long recurrence time.

According to standard theory of binary evolution the
Roche lobe overflow from the companion star is caused by 
loss of angular momentum through gravitational radiation or magnetic
braking or by expansion of the companion during
nuclear evolution off the main sequence. The question which mass transfer
rates are expected in black hole SXTs was already addressed earlier.  
King et al. (1996) considered the transient behaviour in low mass
X-ray binaries based on magnetic braking and/or gravitational
radiation and pointed out that irradiation of the disk
by the central source is important for neutron star binaries but absent
in black hole systems.

The present investigation concerns the long recurrence times.
The theoretical modelling explains the long or even extremely long
recurrence times, but it is not understood why we observe mainly 
systems with very rare outbursts. Could there be any mechanism which 
influences the mass overflow from the companion star in a way that
rare outbursts are favored? We make a suggestion for such a kind of
``self-regulating braking''. We suggest that a circumbinary disk
irradiated in outbursts has such an effect.

\begin{table}

\caption{Black--hole transient sources}

\begin{tabular}{|llllllr|}
\hline
\hline
& & & & & & \\
Source & X-ray&BH mass & comp. & $P_{\rm{orb}}$ &outburst&$\Delta t$
\\
name & Nova& ($M_\odot$) & star & (h)& year& (ys) \\
& & & & & & \\
\hline
& & & & & &\\
J0422+32 &  Per & $>$3.2 & M 2 V & 5.1 & '92 & $>$30 \\
A0620-00 & Mon &$>$7.3 &K 5 V &7.8 &'17,'75 & 58\\
GS2000-25&  Vul& 6-7.5&K 5 V &8.3 &'92 &$>$30 \\
GS1124-68& Mus &$\sim 6$ &K 0-4 V &10.4 &'91 &$>$30 \\
H1705-25& Oph &$\sim$ 6 &K &12.5 &'77  &$>$30 \\
4U1543-47&Lup &2.7-7.5 &A 2 V & 27.0&'71,'83,'92  &$\approx 10$
\\
J1655-40& Sco &7.02$\pm$0.22 &F 3-6 IV &62.7 &'94 &$>$30 \\
GS2023-338& Cyg &8-15.5 &K 0 IV &155.3 &'38,'56,'79,'89&10-20 
\\
& & & & & & \\
\hline
\end{tabular}
\vspace {1cm}
\\
Note: 
Systems established as black hole transients, data for
black hole mass, spectral type of companion star, orbital period and 
outburst year from Charles (1998).

\end{table}

\section{Modelling the outburst cycles of X-ray transients}

The observations for X-ray novae in outburst and quiescence provide
constraints for the theoretical modelling such as an estimate of the amount
of matter accreted during the outburst, values of the accretion rate
in quiescence from the ADAF model based spectral fits, the
radius of the transition from the thin disk to the ADAF
from the maximum velocity width of the
accretion disk H$\alpha$ line (Kepler velocity) (for a review see
Narayan et al. 1998).

The only free parameters for the disk evolution during quiescence are the
mass overflow rate $\dot M$ and the viscosity parameter for the cool state
$\alpha_{cool}$. A further ingredient is the amount of matter and its
distribution in the disk right after the outburst. King \& Ritter
(1998) showed that irradiation can explain the long-lasting
exponential outburst decline observed for X-ray transients. This results
in a practically empty disk after the outburst. Therefore we assume
that only little mass is left in the disk at the beginning of the new
quiescence. In this case the disk evolution in
quiescence and outburst can be studied separately. The accumulation of
mass in the disk during the long quiescence and the question when o
a new outburst is  triggered does not depend on the
foregoing outburst.
Our computer code for the evolution of the disk in quiescence
(Meyer-Hofmeister \& Meyer 1999) includes a variable
inner and outer disk edge (the inner edge determined by
evaporation, the outer edge by conservation
of angular momentum and the 3:1 resonance appropriate for systems with
a low mass ratio (Whitehurst 1988, Lubow 1991)). 

For the modelling of A0620-00 we took the standard viscosity value 
$\alpha_{cool}$= 0.05, used also for dwarf nova outburst modelling.
(We point out that despite
similarities in the outburst cycles of SXTs and WZ Sge stars a very
small value of $\alpha_{cool}$ used for the modelling of the latter
(Meyer-Hofmeister et al. 1998) cannot be adequate here, otherwise the total
amount of matter in the SXT disk would be much too high.) That the modelling
of A620-00 fits the observational constraints makes these
results a promising description of SXT disk evolution. Following the
method used there we also studied the evolution of the disks
for binaries with different black hole mass and different mass
overflow rates. Orbital periods of 4, 8 and 16 hours were
considered. The period
determines the size of the accretion disk. The computations for this
range of parameters describe the variety of outburst behavior in
the observed black hole X-ray binaries. One of the results of
theoretical modelling is that about one half of the mass transferred from the
companion star is accreted on to the black hole during quiescence and
one half during the outburst (wind loss neglected).
The fraction of accumulated mass to transferred mass decreases with
increasing outburst recurrence time, but is surprisingly similar
for different black hole masses. This fact is related to the
evaporation efficiency which increases with black hole mass
(Meyer-Hofmeister \& Meyer 2000).

\begin{figure}
\mbox{}\epsfysize9cm\epsfbox{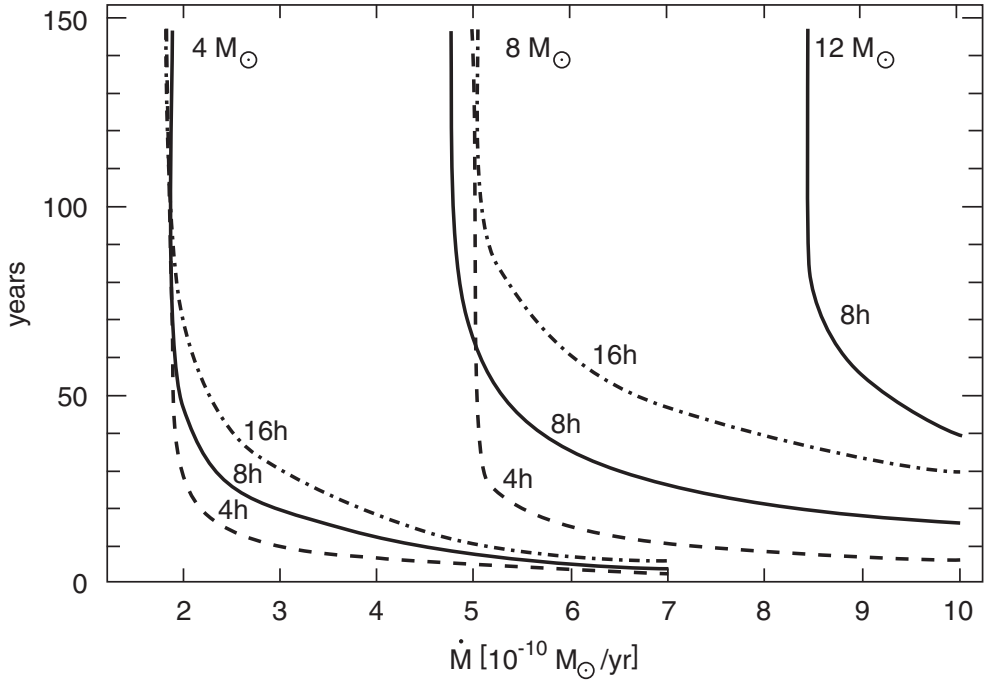}\hfill\mbox{}
\caption{Computed outburst recurrence time for different black
hole masses and orbital periods of the binary as a function of the
mass transfer rate ${\dot M_T}$ from the companion star. Note that the
recurrence time increases steeply with decreasing ${\dot M_T}$ as the
situation of only marginal triggering of
the disk instability is approached. For even lower rates the disk is stable.}
\end{figure}

In Fig.1 we show the computed recurrence times. Given 
a black hole mass and orbital period (which determines the disk size,
larger for longer periods) the outburst recurrence time decreases
with increasing mass transfer rate. This
behavior is expected because the critical surface density at which an
outburst is triggered (for details see Meyer-Hofmeister \& Meyer 1999)
is reached later for a lower transfer rate. For very low rates the matter
accumulation never reaches the critical surface density value and no
outburst occurs. The difficulty to trigger an outburst for relatively
low transfer rates is reflected in the steep increase of recurrence time. 

The dependence of the evaporation on the black hole mass influences the
location of the transition from disk to ADAF and therefore the mass flow 
and accumulation of matter. A larger black hole mass results in more
evaporation and a higher mass transfer rate is needed to trigger an
outburst after a given time interval, for example higher by a factor 2.5
for a disk around a $8M_\odot$ than for a disk around a $4M_\odot$
black hole, as can be seen from Fig. 1. But from Fig. 1 we see also
that for a spread of $\dot M$ we would expect only very few systems with
extremely long and many with shorter recurrence times. The same is true for
a sample of SXTs with different black hole masses.

In Table 1 we give data for systems established as black hole
transient sources. Charles (1998) discusses in detail all constraints and
procedures for the derivation of the black hole masses and gives
ranges of uncertainty. In other compilations for
some systems narrower ranges are given (see also Tanaka \& Shibazaki
1996, Chen et al. 1997,
(Ritter \& Kolb 1998). For three further candidates for black holes
V4641 Sgr, J1858+2239 and MN Vel only few data are available.

For a remarkable number of black hole SXTs only one outburst was
observed. Chen et al. (1997) discussed the properties of X-ray novae, 
the event distribution over 30 years, sky coverage and detection
probability. 
A detection of an optical nova outburst prior to 1975 on archival
plates as for A0620-00 (Nova Mon 1975) and GS2023+338 (Nova Cyg 1989)
vis difficult because of the low appearant luminosity (see discussion
in Romani 1998). But even if an outbursts would have been missed, the
question arises why we find such long recurrence times.
In our theoretical picture this means the mass transfer rates would have to
be the marginal ones, and different for different black hole masses.
This seems implausible if not these rates were
adjusted in some way to cause the long recurrence times.

\section{Magnetic braking, mass overflow rates and resulting
recurrence times}

The problem posed by the observed extremely long recurrence times
brings us to an interesting question: what causes the mass
transfer rates? In Fig. 2 we show the mass transfer rates resulting from
different angular momentum loss mechanisms. If the secondary would be a main
sequence star the rates from magnetic braking would clearly be higher
than the estimates from observation for SXTs (compare Tanaka \&
Shabazaki 1996).
But King et al. (1996) had already argued that the radius
of the secondary star in black hole binaries must be larger than its
main sequence radius if it is to fill its Roche lobe. We therefore
give here the transfer rates for half of the main sequence star mass
(following King et al. 1996). The comparison of rates shows that the
magnetic braking (reduced mass) could just give the rates necessary to
produce outbursts. But if this mechanism determines the transfer rate
we should also find many systems with shorter recurrence time.

\begin{figure}
\mbox{}\epsfysize9.cm\epsfbox{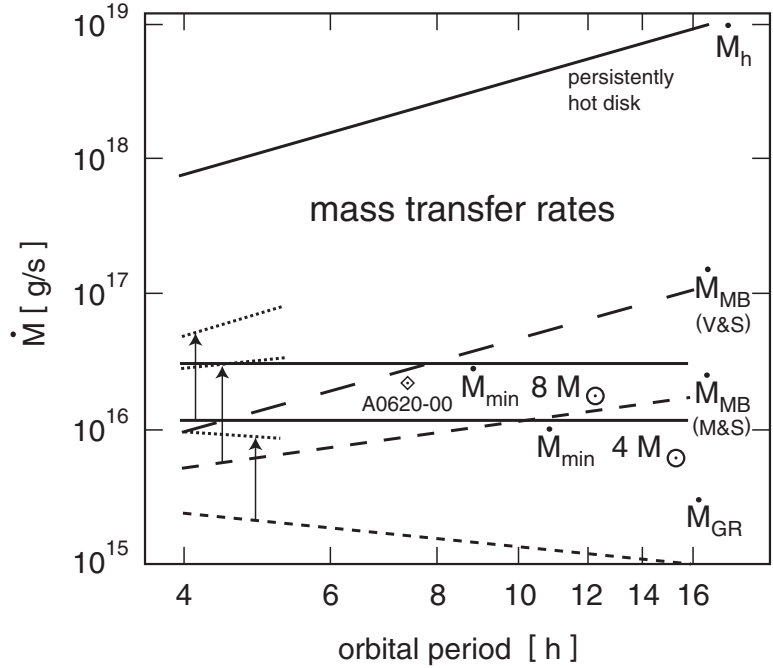}\hfill\mbox{}
\caption{
Mass transfer rates ${\dot M}$ from the companion star due to
different mechanism. Long and short dashed lines: magnetic braking as
suggested by Verbunt \& Zwaan (1981) and Mestel \& Spruit (1987),
very-short-dashed line: braking by gravitational radiation.  (For all
rates is assumed: the black hole mass is 6$M_{\odot}$, the companion
star mass is half of that of a main sequence star for the given
orbital period (that is a slightly evolved star); for a
main sequence star mass the lines would lie higher in the diagram as
indicated by the dotted lines). Upper solid line: limiting mass transfer rate
above which the disk is persistently hot. The two horizontal solid lines
$\dot M_{\rm{min}}$ are the minimal mass transfer rates necessary to
get outbursts for 4 and 8 $M_{\odot}$, practically independent of
orbital period (compare Fig. 1). Black hole mass taken 6$M_{\odot}$.
The position of A620-00 in the diagram is indicated.}
\end{figure}

\section{An mechanism of self-adjusted braking}
\subsection{General properties of a self-adjusting braking mechanism} 
We search for a braking mechanism which establishes the mass transfer rates
so that long recurrence times result, a kind of ``adaptive braking
procedure''. (We then assume that magnetic braking is not the dominant
mechanism). If the outburst should occur every 30 or 50 years, or even
more rarely, such a mechanism must work in the way that the recurrence
time increases if the outbursts tend to appear more often and vice versa.
We here discuss the braking by a circumbinary disk and the effect of
irradiation during the outburst on this disk.

\subsection{The circumbinary disk}
Matter in a circumbinary disk might have remained from an earlier
common envelope
phase when binary stars with an extreme mass ratio evolve into a helium star 
and a low mass companion, or during the expected subsequent collapse
of the helium star into a black hole (for recent work on black hole
formation see e.g. Kalogera 1999).
Fossil disks formed by fallback matter in supernova explosions
have been discussed in connection with accretion on to 
anomalous X-ray pulsars( Perna et al. (2000) and are indicated by the
observation of planets around a neutron star (Wolszczan 1994).

In the evolution of such disks mass is settling inward while angular
momentum is transported outward. At the inner edge mass flow is halted
by the tidal transfer of angular
momentum from the interior binary which thereby is braked. The more
mass piles up at the inner edge of the circumbinary disk the stronger
is the braking of the binary orbit and the stronger
is the induced mass transfer from secondary to primary.

There are now two effects that work in such a system in a way 
to let us see more soft X-ray transients in a phase of marginal mass
transfer than in phases of higher transfer rates.
One is the natural aging of the circumbinary disk, As it receives
angular momentum from the inner binary it spreads viscously to larger and
larger distances and its mass distribution flattens. This relieves the
braking torque at its inner edge and leads to gradually decreasing
inner binary transfer rates. The diffusive ``aging'' of such disks
typically in
power law decay with time (e.g. Meyer \& Meyer-Hofmeister 1989) such
that the duration of each phase is proportional to the age of the
system at this phase. Therefore one finds ten times more systems with about
marginal mass transfer rates than with ten times higher transfer rates.

\subsection{Irradiation}
The other effect is the irradiation of the circumbinary disk by X-rays
during the outbursts of the inner binary (Fig. 3). This results in the
formation of a hot corona with X-ray temperatures. At that distance
from a primary black hole where the inner circumbinary disk finds
itself, of the order of the binary separation, these temperatures are
high enough to drive a wind with considerable mass loss. If outbursts
occur significantly frequently such mass loss
is enough to remove all matter from the inner circumbinary disk which
is continuously settling inward.

\begin{figure}
\plotone{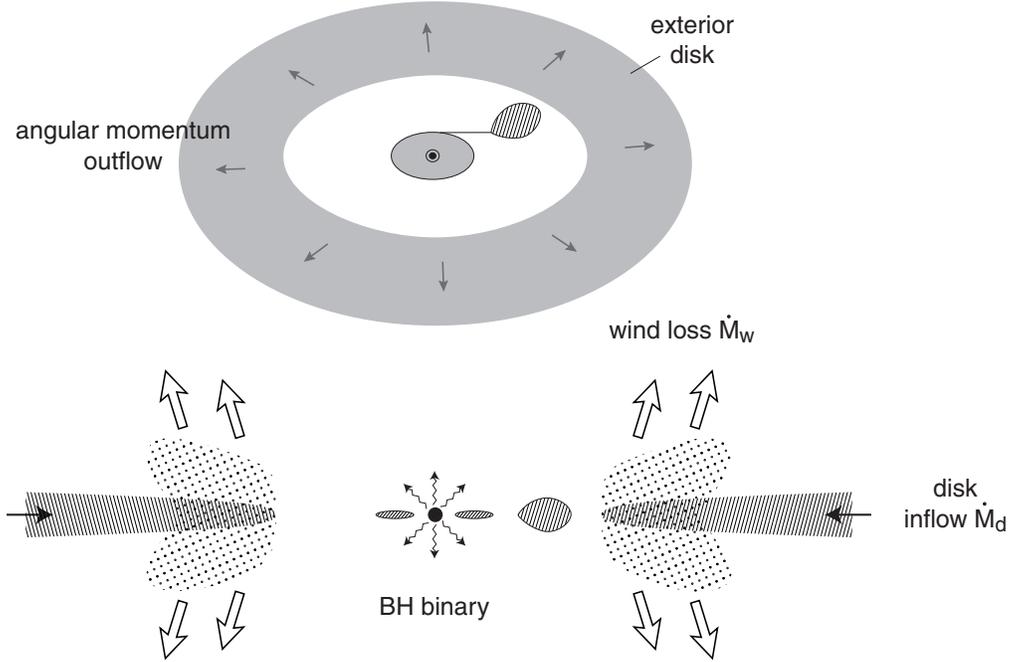}
\caption{Schematic drawing of a circumbinary disk around the black
hole binary, upper drawing. Irradiation caused corona during the
outburst and wind loss are indicated below}
\end{figure}

Whether this mechanism constitutes an efficient self-regulating machine
to bring soft X-ray transients always close to marginal mass transfer
rates and long quiescent intervals depends on the strength of the mass
loss rates during the outbursts. If a large amount of mass is lost
during each outburst infrequent outbursts suffice to keep the braking
rate low. We have performed estimates of the rate of mass loss by wind
using a detailed analysis performed for Her X-1 (Schandl \& Meyer
1994) as a guide line.

Irradiation is reduced by the foreshortened appearance of the flat 
X-ray radiating inner accretion disk and by possible shadowing through
a scattering optically deep coronal layer in the inner region. The
softer X-rays (1 keV) of the SXTs, on the other hand are more
effectively photo-absorbed by metals than the harder ones (10 keV) of
Her X-1 giving more efficient heating and higher pressure in the
corona in comparison.

We have estimated the effects and calculated the wind loss from such
an irradiation caused corona above the inner region of a circumbinary
disk, at a characteristic distance of $10^{11.6}$cm from the inner
black hole system, appropriate e.g. for A0620-00. Integrating over the
X-ray light curve with an initial luminosity of $10^{38}$ erg/s 
exponentially decreasing on a timescale of 80 days we obtain a mass
loss $\Delta M_{\rm w}$ by wind from the circumbinary disk of 
\begin {displaymath} \Delta M_{\rm w}=10^{25.7}{\rm g}  \end{displaymath}
per outburst. If outbursts occur every 50 years this constitutes a
mean mass removal rate of 
\begin {displaymath} \dot M_{\rm w}=10^{16.5} {\rm {g/s}} \end{displaymath}
and correspondingly higher or lower for more or less frequent
outbursts.

It can be shown that the removal of mass at such a rate reduces the braking
efficiency of the circumbinary disk by very closely the amount whose
braking effect would have induced a mass transfer rate from the
secondary to the primary of the same amount.
One can now see in Fig. 2 that this is of the right order of magnitude
to efficiently affect the rate of occurrence of outbursts in these 
soft X-ray transients.

\section{Conclusions}
Our estimates for the rates of mass transfer from the companion star
caused by braking by a circumbinary disk need confirmation by more
accurate calculations. But they indicate that a circumbinary disk
could explain why most of the observed soft X-ray transients appear
to operate at marginal mass transfer rates. With a circumbinary disk
providing most of the orbital braking these systems, as long as they
are visible, will spend most of their time in phases where their
braking without irradiation would produce mass transfer above but not
an order of magnitude above the marginal rate. In these extended
phases however, the irradiation produced wind loss will shift the
actual braking rate very close to the marginal one, as observed.

\vspace{1cm}
\it{This contribution is dedicated to Peter Eggleton on the occasion
of his $60^{th}$ birthday in respect and gratitude for his so
fundamental, rich, and fruitful work.}
\rm

\end{document}